\documentstyle[12pt]{article}
\newcommand{\bq}{\begin{equation}}
\newcommand{\eq}{\end{equation}}
\newcommand{\ba}{\begin{eqnarray}}
\newcommand{\ea}{\end{eqnarray}}

\newcommand{\nobody}{\rule{0ex}{1ex}}

\newcommand{\nobodyfrac}{\frac{\nobody}{\nobody}}
\newcommand{\req}[1]{(\ref{#1})}
\begin{document}
\newtheorem{tab}[table]{Table}
\voffset -1.5cm
\begin{flushleft}
LMU-10/97
\end{flushleft}
\vspace{0.2cm}\hfill\\
\begin{center}
{ \Large \bf Scaling of $Z'$ limits at the Linear Collider}\footnote{
Contribution to the workshop on ``$e^+e^-$ collisions at $TeV$ energies: The
physics potential'', DESY 97-123E}
\vspace{0.2cm}\\ 
A. Leike\\
{\it Ludwigs--Maximilians-Universit\"at, Sektion Physik, Theresienstr. 37,\\
D-80333 M\"unchen, Germany}\\
E-mail: leike@graviton.hep.physik.uni-muenchen.de\\
\end{center}
\hfill\vspace{0.2cm}\\
\noindent
{\small 
$Z'$ exclusion limits and errors of $Z'$ model measurements are
compared for different reactions at future linear colliders.
The influence of the c.m. energy, integrated luminosity, beam polarization and
systematic errors is discussed.
The sensitivity to a $Z'$ depends only on the product $Ls$ and not on
the integrated luminosity and the c.m. energy separately.
}
 \vspace{0.4cm}
%
\section{Introduction} 
%
Extra neutral gauge bosons ($Z'$) are predicted in grand unified
theories (GUT's) with a unification group larger than $SU(5)$.
They are candidates for particles, which could be discovered at the
next linear collider.

In the previous years, extensive studies on the sensitivity of future
$e^+e^-$ colliders to a $Z'$ have been completed \cite{ee}. 
See these references for more details and for related original references. 

It is important to have a simple
understanding how the sensitivity to new physics depends on the collider
parameters, i.e. the c.m. energy $\sqrt{s}$ of the colliding particles, the
integrated luminosity $L$, the beam polarization and the expected
systematic errors of the different observables.
We consider here the sensitivity of different reactions to a $Z'$ with
the help of 
simple analytical formulae predicting the results of time consuming
exact analyses with an acceptable accuracy.\vspace{2mm}

We assume that the interaction of neutral gauge bosons with Standard
Model (SM) fermions is given by the lagrangian
\bq
\label{linkapp}
{\cal L} = 
e A_\beta J^\beta_\gamma + g_1 Z_\beta J^\beta_Z + g_2 Z'_\beta J^\beta_{Z'},
\end{equation}
where the first two terms are SM contributions and the last term is
due to the $Z'$.
The gauge bosons couple through vector and axial vector couplings,
\bq
\label{link}
J^\beta_X=\sum_f\bar{u}_f\gamma^\beta\left[\nobodyfrac\gamma_5 a_f(X)
+ v_f(X) \right] u_f,\ \ \  
X=\gamma,Z,Z'.
\eq

In general, we must admit that the symmetry eigenstates $Z$ and $Z'$
are different from the mass eigenstates $Z_1$ and $Z_2$.
Mass and symmetry eigenstates are linked by a mixing matrix
parametrized by the $ZZ'$ mixing angle $\theta_M$, 
\bq
\label{zzmix}
\left( \begin{array}{c} Z_1 \\ Z_2 \end{array} \right)
=
\left( \begin{array}{rl}  \cos\theta_M & \sin\theta_M \\
                        - \sin\theta_M & \cos\theta_M \end{array} \right)
\left( \begin{array}{c} Z \\ Z' \end{array} \right).
\eq
The masses and the widths of the mass and symmetry eigenstates are
denoted as $M_i,\Gamma_i,\ i=1,2,Z,Z'$.
The mixing must be small because the experiments at LEP\,1
and SLC agree precisely with the SM.

We assume no mixing between the SM fermions and the new
fermions present in a GUT.
We assume that all new particles in the GUT are heavier than the $Z'$.
See \cite{e6} for a review on $Z'$ models and further references.

The experimental error of an observable $O$ consists of a systematic
$\Delta^{syst} O$ and a statistical $\Delta^{stat}O$ contribution.
We add both contributions in quadrature,
\bq
\Delta O = \sqrt{(\Delta^{stat}O)^2+(\Delta^{syst}O)^2}
= \Delta^{stat}O\cdot\sqrt{1+r^2},\ \ \ r=\Delta^{syst}O/\Delta^{stat}O.
\eq 

One has to distinguish between $Z'$ {\it exclusion
limits} and $Z'$ {\it model measurements}.
Exclusion limits will be obtained if there are no
deviations from the SM.
All present bounds on $Z'$ theories are examples of exclusion limits.
Model measurements will be possible if there are deviations from the SM
predictions compatible with theories containing a $Z'$.\vspace{2mm} 

Lepton colliders have the advantage of a clean environment. 
Highly polarized electron beams are available.
We have several reactions, which are sensitive to extra $Z$ bosons.

{\it Fermion pair production} allows for a measurement of a large number of
different observables.
All couplings of the $Z'$ to charged SM fermions can be constrained
separately.
This is a unique property of this reaction.

{\it Bhabha} and {\it M{\o}ller scattering} have large event rates.
In addition, M{\o}ller scattering profits from two highly
polarized electron beams. 
Of course, these reactions are sensitive to gauge boson couplings to
electrons only. 

{\it W pair production} is very sensitive to $ZZ'$ mixing.
This sensitivity is enhanced for large energies because additional amplitudes
destroy the gauge cancellation present in the SM.

All {\it other reactions} in  $e^+e^-$ or $e^-e^-$ collisions can not
add useful information on extra neutral gauge bosons.
\section{$e^+e^-\rightarrow f\bar f$}
Off-resonance fermion pair production is almost insensitive to $ZZ'$
mixing.
We therefore set $\theta_M=0$ in this section.

As has been proven by the LEP and SLC experiments, different
fermions as $e,\mu,\tau,c,b$ can be tagged in the final state. 
The tagging of top quarks is expected at the linear collider. 
The polarization of $\tau'$s and likely of top quarks can be measured.

Fermion pair production is sensitive to a $Z'$ at energies far below
the $Z'$ peak. 
A $Z'$ modifies cross sections and asymmetries due to 
interferences of the $Z'$ amplitude with the SM amplitudes.
With measurements below the $Z'$ resonance, only the ratios
$a_f(Z')/M_{Z'}$ and $v_f(Z')/M_{Z'}$ can be constrained 
and not the couplings and the mass separately, unless very
high luminosities are available and the systematic errors are very small.
\subsection{Exclusion Limits}
An agreement of future measurements of an observable $O$ with the SM
prediction proves the $Z'$ to be heavy, 
\bq
\label{offres}
M_{Z'}>M_{Z'}^{lim}
\approx\frac{g_2}{g_1}\sqrt{s\frac{O}{\Delta O}}
\approx 2.8\,TeV\cdot\frac{g_2}{g_1}
\left[\frac{Ls}{1+r^2}\frac{fb}{TeV^2}\right]^{1/4}.
\eq
The estimate \req{offres} assumes that $\frac{g_2\Delta O}{g_1O}\gg 1$.
We have $g_2/g_1=\sqrt{\frac{5}{3}}\sin\theta_W\approx 0.62$ in the
$E_6$ GUT. 
The scaling of the error with $s$ and $L$ ,
$\frac{\Delta O}{O}\approx\sqrt{(1+r^2)/N}\sim\sqrt{(1+r^2)s/L}$,
is taken into account in the last step of the estimate \req{offres}.
We further assume that all couplings of the $Z'$ to fermions are fixed
by the GUT. 
Then, many observables with leptons and hadrons in the final state
contribute to $M_{Z'}^{lim}$.
If there are no model assumptions on the $Z'$ couplings to quarks,
only observables with leptons in the final state will contribute.
Then, the factor $2.8\,TeV$ has to be replaced by 1.9\,TeV.

$M_{Z'}^{lim}$ depends on the fourth root of the experimental error.
The dependence on the systematic error will be suppressed if
it is not too large.
Suppose that an analysis gives certain exclusion limits $M_{Z'}^{lim}$
without systematic errors.
What changes are expected after the inclusion of systematic errors
assuming $r=1$?
The estimate \req{offres} predicts 
$M_{Z'}^{lim}\rightarrow M_{Z'}^{lim}/\sqrt[4]{2}$,
which is a reduction by 16\% only.

%
\begin{table}[tbh]
\begin{center}
\begin{tabular}{lrrrr}\hline
 &$\chi$ &$\psi$ &$\eta$ &$LR$ \rule[-2ex]{0ex}{5ex}\\ 
\hline
$M_{Z'}^{lim}$ \ stat. & 3.1 & 1.8 & 1.9 & 3.8 \\
$M_{Z'}^{lim}$ +syst. & 2.8 & 1.6 & 1.7 & 3.2 \\
$P_V^l$ \ stat. & 2.00$\pm$ 0.11 & 0.00$\pm$ 0.064 & -3.00$^{+0.53}_{-0.85}$ &
                         -0.148$^{+0.020}_{-0.024}$ \\
$P_V^l$ +syst.& 2.00$\pm$ 0.15 &
                          0.00$\pm$ 0.13 &
                         -3.00$^{+0.73}_{-1.55}$  & 
                         -0.148$^{+0.023}_{-0.026}$  \\
$P_L^b$ \ stat  & -0.500$\pm$ 0.018 &
                           0.500$\pm$ 0.035 &
                           2.00$^{+0.33}_{-0.31}$ &
                           -0.143$\pm$0.033  \\
$P_L^b$ +syst.& -0.500$\pm$ 0.070 &
                           0.500$\pm$ 0.130 &
                           2.00$^{+0.64}_{-0.62}$ &
                           -0.143$\pm$0.066  \\
\hline
\end{tabular}\medskip
\end{center}
{\small\it  \begin{tab}\label{zplimlc} The lower bound on $Z'$ masses
$M_{Z'}^{lim}$ in TeV excluded by $e^+e^-\rightarrow f\bar f$ at
$\sqrt{s}=0.5\,TeV$ and $L=20\,fb^{-1}$ (first two rows).
The $Z'$ coupling combinations $P_V^l=v'_l/a'_l$ and
$P_L^b=(v'_b+a'_b)/(2a'_b)$ and their 1-$\sigma$ errors (last four
rows) are derived under the same conditions as the exclusion 
limits but assuming $M_{Z'}=1\,TeV$.
The $\chi,\psi,\eta$ and $LR$ models are the same as in the Particle
Data Book.
The numbers are given with and without systematic errors.
They are taken from reference \cite{lmu0296}.
\end{tab}} \end{table}
%

Let us confront these findings with the numbers quoted in table~\ref{zplimlc}.
They include all SM corrections. 
The systematic errors included for observables with leptons in
the final state are roughly as large as their statistical
errors, i.e. $r\approx 1$. 
The systematic errors of observables with $b$ quarks in the final
state are roughly 4 times as large as the statistical errors,
i.e. $r\approx 4$.

We see that the predicted reduction of $M_{Z'}^{lim}$ by 16\% is
reproduced by the numbers in the first two rows of table~\ref{zplimlc}.
Although $M_{Z'}^{lim}$ is defined by hadronic and leptonic observables,
hadronic observables with large systematic errors don't spoil the
estimate \req{offres} because their contribution to $M_{Z'}^{lim}$
decreases in that case.

Polarized beams give almost no improvements to $Z'$ exclusion limits
\cite{lmu0296}. 
\subsection{Model Measurements}
Suppose that there exists a $Z'$ with $M_{Z'} <M_{Z'}^{lim}$.
Then, a measurement of the ratios $a_f(Z')/M_{Z'}$ and
$v_f(Z')/M_{Z'}$  is possible. 
Alternatively, one can measure the $Z'$ mass for fixed couplings or the
coupling strength for a fixed $Z'$ mass, which could be known
from hadron collisions.
Considerations \cite{habil} similar to the previous section give an
estimate of the {\it errors} of such measurements as
\bq
\label{epemmeas}
\frac{\Delta M_{Z'}}{M_{Z'}},\ \frac{\Delta g_2}{g_2} 
\approx\frac{1}{2}\left(\frac{M_{Z'}}{M_{Z'}^{lim}}\right)^2
\approx c_f\cdot\frac{g_1^2}{g_2^2}
\frac{M_{Z'}^2}{TeV^2}\left[\frac{1+r^2}{Ls}\frac{TeV^2}{fb}\right]^{1/2}
\sim\left[\frac{1+r^2}{Ls}\right]^{1/2}
\eq
with $c_f\approx 0.063$ for leptons in the final state.
Model measurements depend on the square root of the experimental errors.
Another important difference to the exclusion limit \req{offres} is
that the couplings of the $Z'$ to leptons are measured by
observables with leptons in the final state only, while the 
couplings of the $Z'$ to $b$-quarks are measured by
observables with $b$-quarks in the final state only.

In particular, the estimate \req{epemmeas} predicts (under the
assumptions of the analysis \cite{lmu0296}) that the 
errors of measurements of the $Z'$ couplings to leptons ($b$-quarks)
change as $\Delta P_V^l\rightarrow \Delta P_V^l\sqrt{2}$ ($\Delta
P_L^b\rightarrow\Delta P_L^b\sqrt{17}$) after the inclusion of the
systematic errors. 
Of course, these predictions are only rough
approximations because they ignore details of the $Z'$ models and
differences of the ratio $\Delta^{syst}O/\Delta^{stat}O$ for the various
observables entering the analysis.
Nevertheless, they reproduce the main tendency of the last four rows in
table~\ref{zplimlc}. 
The estimate \req{epemmeas} explains why $Z'$ model measurements are
much more sensitive to systematic errors than $Z'$ exclusion limits.

We remark that the scalings \req{offres} and \req{epemmeas} depend on
the product $Ls$ only and not on $s$ and $L$ separately.
The product $Ls$ is the ``currency'', in which we have to pay for a
$Z'$ search.
The estimate \req{epemmeas} relates the ``price'' $(Ls)_{det}$ one has
to pay for the detection of a $Z'$ of a certain model to the ``price''
$(Ls)_\varepsilon$ of a
model measurement of the same model by the same observables at the
same confidence level with the accuracy $\varepsilon$,
\bq
\label{measexcl}
(Ls)_\varepsilon\approx\frac{1}{4\varepsilon^2}\cdot(Ls)_{det}
\eq

One can go one step further and try to determine the couplings and the
mass of the $Z'$ separately by a fit to the line shape below the $Z'$
resonance.
Such an measurement is proposed in reference \cite{rizzoi}. 
Here is the list of ``prices'' for an investigation
of $Z'=\chi,\ M_\chi=1.6\,TeV,\ 95\% CL.,\ (s<M_2^2,)$:

\begin{tabular}{ll}
Detection: & $Ls\approx 0.5TeV^2/fb$\\
Measurement of $a_e(Z')/M_{Z'}$ and $v_e(Z')/M_{Z'}$ with 15\% error: 
& $Ls\approx 8\,TeV^2/fb$\\
Measurement of $a_e(Z'), v_e(Z')$ and $M_{Z'}^2$ separately with 15\% error: 
& $Ls\approx 260\,TeV^2/fb$
\end{tabular}

The first number is obtained from the first row of table~\ref{zplimlc}
using the scaling \req{offres}.
The second number is obtained from $M_{Z'}^{lim}$ calculated from
leptonic observables only \cite{lmu0296} and scaling \req{epemmeas}.
The third number is calculated from reference \cite{rizzoi}. 
We see that the prices for the detection and the two proposed
measurements are very different.

In contrast to exclusion limits, the electron polarization is very
important for a $Z'$ model measurement. 
It reduces the four-fold sign ambiguity of a $Z'$ coupling measurement
to a two-fold ambiguity \cite{lmu0296}. 
This is a qualitative effect, which cannot by
``bought'' by an increase of $Ls$ in an unpolarized measurement.
Polarization of both beams would not give a further improvement.
\section{$e^+e^-\rightarrow e^+e^-$ and $e^-e^-\rightarrow e^-e^-$}
Bhabha and M{\o}ller scattering are as insensitive to $ZZ'$ mixing as
off-resonance fermion pair production. 
We therefore set the $ZZ'$ mixing angle to zero in this section.
$Z'$ constraints arise by the same mechanisms as in fermion pair
production.
A difference occurs due to the gauge boson exchange in the $t$ and
$u$ channels, which leads to a very singular angular distribution.
Most of the scattered particles are near the beam pipe due to photon
exchange. 
Therefore, the total cross sections are rather insensitive to $Z'$
effects.
Angular distributions are sensitive to $Z'$ exchange.
\subsection{Exclusion Limits}
The exclusion limits from Bhabha and M{\o}ller scattering scale as in
fermion pair production, 
\bq
\label{offres2}
M_{Z'}>M_{Z'}^{lim}
\approx 1.9\,TeV\cdot\frac{g_2}{g_1}
\left[\frac{Ls}{1+r^2}\frac{fb}{TeV^2}\right]^{1/4}.
\eq
$g_1$ and $g_2$ are here the coupling strengths to electrons.
The numerical factor $1.9\,TeV$ is obtained from reference \cite{moeller2}. 
The exclusion limit from  M{\o}ller scattering is comparable to that
from fermion pair production with leptons in the final state.
This agrees with the results of reference \cite{9609248}.

In contrast to fermion pair production, the polarization of both beams
is important for the exclusion limits in M{\o}ller scattering \cite{moeller1}. 
However, it improves the limits only quantitatively.
An equivalent gain could be ``bought'' by unpolarized beams with
approximately 6 times higher $Ls$.

As M{\o}ller scattering, Bhabha scattering has a definite final state.
With an unpolarized positron beam, the constraints from Bhabha
scattering are weaker than those from M{\o}ller scattering \cite{moeller3}. 
However, Bhabha scattering does not need a special option of the
linear collider.
A polarized positron beam is expected to improve the $Z'$ constraints from 
Bhabha scattering.
\subsection{Model Measurements}
Bhabha and M{\o}ller scattering can only constrain the couplings of the
$Z'$ to electrons.
Fermion pair production and M{\o}ller scattering are complementary in
model measurements involving these couplings \cite{moeller2}.
Fermion pair production reduces the sign ambiguities
present in the measurements with Bhabha and M{\o}ller scattering.
The estimate \req{epemmeas} for leptons in the final state applies.
\section{$e^+e^-\rightarrow W^+W^-$}
The individual interferences between the different amplitudes to $W$
pair production rise like $s$. 
The two leading powers in $s$ cancel due to gauge cancellations making
the total cross section proportional to $\ln s/s$.
Due to the $SU_L(2)$ gauge invariance, a $Z'$ can couple to a $W$ pair
only in presence of a non-zero $ZZ'$ mixing.
These additional contributions to $W$ pair production destroy the
cancellation mechanism leading to a huge enhancement of the
sensitivity to a $Z'$.

A $Z'$  signal in $W$ pair production can be absorbed \cite{npb429} in ($s$
dependent) anomalous couplings of the photon and the $Z$ boson
to the $W$ pair.
$W$ pair production can constrain the combinations $a_e(Z')\theta_M$
and $v_e(Z')\theta_M$ and not the couplings of the $Z'$ to electrons
and the $ZZ'$ mixing angle separately \cite{pankovnew}. 
\subsection{Exclusion Limits}
Polarized electron beams are necessary to constrain {\it all} $Z'$ models.
Without polarized beams, an infinite band in the $a_e(Z')\theta_M,
v_e(Z')\theta_M$ plane would be allowed \cite{pankovnew}.
Polarized positron beams would not give a further improvement.
Assuming $v_e(Z')\approx v_e(Z)$ and $a_e(Z')\approx a_e(Z)$, one can
estimate the sensitivity to $\theta_M$ as \cite{habil} 
\bq
\label{wwtmest}
|\theta_M|<3.4\cdot\frac{\Delta\sigma_T}{\sigma_T}\cdot\frac{M_Z^2}{s}
\frac{g_1}{g_2}
\Re e\left|1-\frac{s-M_Z^2}{s-M_2^2+i\Gamma_2M_2}\right|^{-1}
\sim\left[\frac{1+r^2}{Ls}\right]^{1/2},
\eq
where $\sigma_T$ is the total cross section.
$\Delta\sigma_T$ in the estimate is dominated by systematic errors.
Again, the best $Z'$ constraints are obtained by fits to the angular
distribution. 
The factor $\Re e(\dots)$ is one for $s\ll M_2^2$ and approximately
$2\Gamma_2/M_2$ for $\sqrt{s}= M_2\pm\Gamma_2/2$.
The exclusion limit is as sensitive to systematic errors as
the model measurement in fermion pair production.
Assuming that future colliders will measure $W$ pair production with
a certain relative error, we see that the sensitivity
to $\theta_M$ is enhanced by the factor $M_Z^2/s$.
It is enhanced by the additional factor $2\Gamma_2/M_2$ in measurements
near the $Z_2$ peak \cite{pankovnew}.
These enhancement factors overcompensate the effects of the larger
statistics in fermion pair production.
As a result, the sensitivity of $W$ pair production to
$\theta_M$ is much larger than that of fermion pair production at
the $Z_2$ peak \cite{pankov}.
\subsection{Model Measurements}
One could try a $Z'$ model measurement in the case of a non-zero $ZZ'$ mixing.
The error of such a measurement is expected to scale with $L,s$ and $r$
as \req{wwtmest}.
\section{Conclusion}
A linear collider allows to study $Z'$ effects in different reactions.
Off resonance fermion pair production, Bhabha and M{\o}ller scattering
constrain ratios of $Z'$ couplings and the $Z'$ mass.
They are insensitive to $ZZ'$ mixing.
For a fixed $Z'$ model, the mass exclusion limits from fermion pair
production are better than those from Bhabha or M{\o}ller scattering.
If there are no model assumptions linking the couplings of the $Z'$ to
leptons and quarks, the exclusion limits from fermion pair
production, Bhabha and M{\o}ller scattering will be comparable.
Fermion pair production, Bhabha and M{\o}ller scattering are
complementary in a model measurement involving the
couplings of the $Z'$ to electrons.  

$W$ pair production is very sensitive to $ZZ'$ mixing.
In a fixed model, the $ZZ'$ mixing angle and the $Z'$ mass are linked.
Then, the $Z'$ constraints from all considered reactions can be
compared.

Polarized electron beams give important improvements to $Z'$ exclusion limits
except fermion pair production.
They are very important in all reactions for model measurements.
Polarization of the positron beam gives almost no improvement for a $Z'$ search.

In all reactions, the sensitivity to a $Z'$ depends on the
combination $Ls/(1+r^2)$ only, and not on the integrated luminosity, the
c.m. energy squared and the ratio of the systematic and statistical
errors separately.  
$Z'$ exclusion limits are always less sensitive
to systematic errors than $Z'$ model measurements.

\end{document}